\documentclass[12pt]{article}
\usepackage{amssymb,amsmath,amscd}
\newcommand{\lto}{\longrightarrow}

\begin{document}
\begin{center}
{\Large \bf Noncommutative Riemann Conditions}

\vspace{12mm} Eunsang Kim${}^{a,}$\footnote{eskim@ihanyang.ac.kr},
\ and \; Hoil Kim${}^{b,}$\footnote{hikim@knu.ac.kr} \

\vspace{5mm} ${}^a${\it
Department of Applied Mathematics,\\ Hanyang University, Ansan Kyunggi-do 425-791, Korea}\\
${}^b${\it Department of Mathematics,\\
Kyungpook  National University, Taegu 702-701, Korea}\\

\vspace{12mm}

\end{center}

\begin{center}
 {\bf Abstract}

\vspace{5mm}

\parbox{125mm}{In this paper we study the holomorphic bundles over
a noncommutative complex torus. We define a noncommutative abelian
variety as a kind of deformation of abelian variety and we show
that for a restricted deformation parameter,  one can define a
noncommutative abelian variety. Also, along the cohomological
deformation, we  discuss the noncommutative analogue of usual
Riemann conditions. This will be done by using the real
cohomologies instead of the rational ones.

} \vfill
\end{center}
\setcounter{footnote}{0}

\pagebreak

\section{Introduction}
The noncommutative tori is known to be the most accessible
examples of noncommutative geometry developed by A. Connes
\cite{Connes94B}. It also provides the best example in
applications of noncommutative geometry to string/M-theory which
was initiated in \cite{CDS}. Analogously the geometry and gauge
theory of noncommutative torus has been explicitly studied in many
papers, such as \cite{CDS}, \cite{CoR}, \cite{Ri1}, \cite{Sch98}.
Rather recently, complex geometry of noncommutative torus was
developed by A. Schwarz \cite{Sch01} and  some detailed analysis
were made in \cite{DiSch} for the two-dimensional case (see also
\cite{KL} for four-dimensional ones). It provided a basic result
for subsequent papers \cite{PolSch}, \cite{Kaj} and \cite{Pol}
which study the Kontevisch's homological mirror symmetry
conjecture \cite{Ko} on the two-dimensional noncommutative tori.
One of our main underlying motivations is to study the mirror
symmetry on the higher dimensional cases and the goal of this
paper is to set up a very first step for that study.

Our main interest in this paper is to compare complex geometries
of both noncommutative  and ordinary complex torus, focused on the
four-dimensional case. As discussed in \cite{GH}, the ordinary
Riemann conditions are to determine whether a complex torus is an
abelian variety or not. We show that the definition of holomorphic
structures given in \cite{Sch01} is a noncommutative analogue of
Riemann conditions, which determines the existence of theta
vectors. Since noncommutative torus is defined as a deformation
quantization of an ordinary torus \cite{Ri0}, we may regard a
holomorphic bundle over a noncommutative complex torus as a
deformation of a bundle over a commutative complex torus. Compared
with other examples of noncommutative geometry, the Chern
character of noncommutative torus is easily understood because it
takes the values in the real cohomological group of the
commutative torus \cite{Ri1}. The Chern character can be obtained
from the ordinary Chern character via deformation parameter. Our
deformed bundle will be given in terms of constant curvature
connections such that the curvature matrix reflects the Chern
character deformation. Such a bundle equipped with a constant
curvature connection is easily seen as a deformation of a
projectively flat bundle over a complex torus. From this point of
view, we will define a noncommutative abelian variety as a
deformation of abelian variety and we find that such a
noncommutative abelian variety can be defined for a restricted
deformation parameter. We will also study noncommutative analogue
of ordinary Riemann conditions given in \cite{GH}. This will be
done by using the real cohomologies instead of the rational ones,
considering the usual Riemann conditions are obtained by
compairing the rational structure and the complex structure of the
cohomology of a torus.

This paper is organized as follows. In Section 2, we review a
general concepts on noncommutative complex tori defined in
\cite{Sch01} and we define a deformed bundle using the Chern
character deformation. By regarding an abelian variety as a pair
of complex torus and a line bundle on it, we define a
noncommutative abelian variety, which is a deformation of abelian
variety. In Section 3, we first find an explicit solutions for the
theta vectors on a four-dimensional noncommutative complex torus.
We also discuss the analogue the definition of holomorphic bundles
in two different ways depending on the complex structures to
define a noncommutative complex torus. We conclude in Section 4.

\section{Preliminaries}

Let $\mathbb{T}^d_\theta$, $d=2g$,  be a $d$-dimensional
noncommutative torus. It is generated by $d$-unitaries
$U_1,\cdots,U_d$ such that
\begin{equation}\label{1}
U_iU_j=\exp(2\pi i\theta_{ij})U_jU_i,
\end{equation}
where $\theta=(\theta_{ij})$ is a real $d\times d$ skew-symmetric
matrix. The relation (\ref{1}) defines the presentation of the
involutive algebra
\[A_\theta^d=\left\{
\sum_{(n_1,\cdots,n_d)\in\mathbb{Z}^d}a_{n_1,\cdots,n_d}U_1^{n_1}\cdots
U_d^{n_d}\mid a_{n_1,\cdots,n_d}\in
\mathcal{S}(\mathbb{Z}^d)\right\}\] where the coefficient function
$(n_1,\cdots,n_d)\mapsto a_{n_1,\cdots,n_d}$ rapidly decays at
infinity. By definition, the algebra $A_\theta^d$ is the algebra
of smooth functions on $\mathbb{T}^d_\theta$ and the bundles over
$\mathbb{T}^d_\theta$ correspond to finitely generated projective
(left) $A_\theta^d$-modules.

The ordinary torus $\mathbb{T}^d$ acts on the algebra $A_\theta^d$
({\it cf.} \cite{Ri1}) and the infinitesimal form of the action of
$\mathbb{T}^d$ on $A_\theta^d$ defines a Lie algebra homomorphism
\[\delta:L\lto \text{ Der}(A^d_\theta),\]
where $L\cong \mathbb{R}^d$ is the Lie algebra of $\mathbb{T}^d$
and $\text{ Der}(A^d_\theta)$ denotes the Lie algebra of
derivations of $A_\theta^d$. Thus for $x\in L$,
\[\delta_x(uv)=\delta_x(u)v+u\delta_x(v), \ \ u,v\in A_\theta^d.\]
For a basis $\{\lambda_i\}_{i=1,\cdots,2g}$ for $L$, set
$\delta_{\lambda_i}:=\delta_i$. Then
\[\delta_j(U_j)=2\pi iU_j, \ \
\delta_i(U_j)=0 \ \ \text{ for  } i\ne j,\] or
\[\delta_j\left(\sum_{\mathbf{n}=(n_1,n_2,\cdots,n_d)\in\mathbb{Z}^d}a_{\mathbf{n}}U_1^{n_1}\cdots
U_d^{n_d}\right)=\sum_{\mathbf{n}\in\mathbb{Z}^d}2\pi in_j
a_\mathbf{n}U_1^{n_1}\cdots U_d^{n_d}.\] Note that the Lie algebra
$\text{ Der}(A^d_\theta)$ plays the role of tangent bundle to
$\mathbb{T}^d_\theta$.

Since $\mathbb{T}^d$ is commutative, the cohomology group
$H^*(\mathbb{T}^d,\mathbb{R})$ can be identified with the exterior
algebra $\wedge^\bullet L^*$, where $L^*$ is the dual vector space
of $L$. Let $(x_1,\cdots,x_{2g})$ be the dual real coordinates on
$L$ and $dx_1,\cdots,dx_{2g}$ the corresponding 1-forms on $T^d$.
Then
\begin{align}\label{coho}
H^*(\mathbb{T}^d,\mathbb{R})=\mathbb{R}\{dx_I\}_I, \ \ \ \ \
H^*(\mathbb{T}^d,\mathbb{Z})=\mathbb{Z}\{dx_I\}_I.
\end{align}

A complex structure on $L$ can be given in the form
\[L\oplus iL\cong L^{1,0}\oplus L^{0,1}=V\oplus\bar V, \ \ V\cong
\mathbb{C}^g.\]
\[L=\mathbb{R}^d=\mathbb{R}^{2g}\cong\mathbb{C}^g:=V.\]
Thus we may consider $\mathbb{T}^d$ as a complex torus
$\mathbb{T}^d=V/\Lambda$, where $\Lambda$ is a lattice of maximal
rank $2g$ which is defined by the kernel of the exponential map
$\exp:V\lto \mathbb{T}^d$.  Let $\lambda_1,\cdots,\lambda_{2g}$ be
an integral basis for $\Lambda$, which will also be a basis for
the real vector space $L$ and let $e_1,\cdots,e_g$ be a complex
basis for $V$. We take the period matrix of $\Lambda\subset V$ to
be a $g\times d$ complex matrix $\mho=(\mho_{\alpha i})$ such that
\[\lambda_i=\sum_\alpha \mho_{\alpha i}e_\alpha.\]
Let $z=(z_1,\cdots,z_g)$ be Euclidean coordinates on $V$ and let
$\{dz_1,\cdots,dz_g\}$ and

\noindent$\{d\bar z_1,\cdots,d\bar z_g\}$ be corresponding 1-forms
on $\mathbb{T}^d$. Then
\begin{align*}
dz_\alpha&=\sum_i\mho_{\alpha i}dx_i\\
d\bar z_\alpha&=\sum_i\bar\mho_{\alpha i}dx_i
\end{align*}
so that the matrix
$\begin{pmatrix}\mho\\\overline{\mho}\end{pmatrix}$ gives the
change of basis from $\{dx_i\}$ to $\{dz_\alpha,d\bar z_\alpha\}$.
We can also identify
\[H^*(\mathbb{T}^d,\mathbb{C})==\wedge^\bullet V^*\otimes \wedge^\bullet
\overline{V^*}=\mathbb{C}\{dz_I\wedge d\bar z_J\}_{I,J}.\]

In the case of complex tori, the Kodaira's embedding theorem is
that the complex torus $\mathbb{T}^d=V/\Lambda$ is an abelian
variety if and only if there is a closed, positive $(1,1)$-form
$\omega$ whose cohomology class $[\omega]$ is rational, so that
$[\omega]\in H^{1,1}(\mathbb{T}^d)\cap
H^2(\mathbb{T}^d,\mathbb{Z})$. We will understand an abelian
variety as a pair $(\mathbb{T}^d,\mathsf{L})$, where
$\mathbb{T}^d$ is a complex torus and $\mathsf{L}$ is a line
bundle whose first Chern class is in $[\omega]\in
H^{1,1}(\mathbb{T}^d)\cap H^2(\mathbb{T}^d,\mathbb{Z})$. In terms
of the period matrix $\mho$, the Riemann condition II in \cite{GH}
is stated as follows: The complex torus $\mathbb{T}^d=V/\Lambda$
is an abelian variety if and only if there exists an integral,
skew-symmetric matrix $q$ such that
\[\mho q^{-1}\mho^t=0, \ \ \ \ -i\mho q^{-1}\overline{\mho^t}>0.\]

For any integral 2-form on $\mathbb{T}^d$, we can find an integral
basis $\lambda_1,\cdots,\lambda_{2g}$ for $\Lambda$ such that
\[\omega=\sum_{i=1}^gm_i dx_i\wedge dx_{g+i}, \ \ \
m_i\in\mathbb{Z}.\] By taking a basis for $V$ the vectors
\[e_\alpha=m^{-1}_\alpha\lambda_\alpha, \ \ \ \alpha=1,\cdots, g\]
the period matrix $\mho$ will be of the form
\[\mho=\begin{pmatrix}M&Z\end{pmatrix}=\begin{pmatrix}
m_1&0&\cdots &0&\mathcal{Z}_{11}&\cdots&\mathcal{Z}_{1g}\\0&m_2&\cdots&0&\cdot&\cdots&\cdot\\
\cdot&\cdot&\cdots&\cdot&\cdot&\cdots&\cdot\\
0&\cdots&0&m_g&\mathcal{Z}_{g1}&\cdots&\mathcal{Z}_{gg}
\end{pmatrix}.\]
The Riemann condition II implies the Riemann condition III in
\cite{GH}: the complex torus $\mathbb{T}^d$ is an abelian variety
if and only if the $g\times g$ complex matrix $\mathcal{Z}$ is
symmetric and $\text{Im }Z$ is positive definite.

The period matrix $\mho=(\mho_{\alpha i})$ of $\Lambda$ in $V$
gives the change of basis $\{dx_i\}$ to $\{dz_\alpha\}$. On the
other hand the Lie algebra of derivations of $A^d_\theta$ is
spanned by the derivations
$\delta_{\lambda_1},\cdots,\delta_{\lambda_d}$ and since the
derivations $\delta_{\lambda_i}$ correspond to vector fields
$\frac{\partial}{\partial x_i}$, we will have to use
dual-change-of-basis matrix
$\begin{pmatrix}\Omega\\\overline{\Omega}\end{pmatrix}$ of
$\begin{pmatrix}\mho\\\overline{\mho}\end{pmatrix}$ which gives
the change of basis from $\{\frac{\partial}{\partial x_i}\}$ to
$\{\frac{\partial}{\partial z_\alpha},\frac{\partial}{\partial
\bar z_\alpha}\}$. Those two matrices are related by
\[\begin{pmatrix}\mho\\\overline{\mho}\end{pmatrix}\cdot
\begin{pmatrix}\Omega\\\overline{\Omega}\end{pmatrix}^t=\text{Id}_d\]
so that we can identify the matrix
$\begin{pmatrix}\Omega\\\overline{\Omega}\end{pmatrix}^t$ with the
matrix which changes the  basis from $\{dz_\alpha,d\bar
z_\alpha\}$ to $\{dx_i\}$. Now the Riemann condition I in
\cite{GH} is given in terms of dual period matrix $\Omega$: The
complex torus $\mathbb{T}^d=V/\Lambda$ is an abelian variety if
and only if there exists an integral, skew-symmetric matrix $q$
such that
\[\Omega q\Omega^t=0, \ \ \ \ -i\Omega q\overline{\Omega^t}>0.\]

We now define a noncommutative complex torus using the matrix
$\Omega=(\omega_{\alpha i})$. Let
\[\delta_{\Omega,\alpha}=\sum_{i}\omega_{\alpha i}\delta_i, \ \ \
\alpha=1,\cdots, g.\] Then
\begin{align*}\delta_{\Omega,\alpha}\left(\sum_{\mathbf{n}\in\mathbb{Z}^d}
a_{n_1,\cdots,n_d}U_1^{n_1}U_2^{n_2}\cdots, U_d^{n_d}\right)&=
\sum_{i}\omega_{\alpha
i}\delta_i\left(\sum_{\mathbf{n}\in\mathbb{Z}^d}
a_{n_1,\cdots,n_d}U_1^{n_1}U_2^{n_2}\cdots, U_d^{n_d}\right)\\&=
\sum_{\mathbf{n}\in\mathbb{Z}^d}2\pi i(\sum_{i}\omega_{\alpha
i}n_i) a_{n_1,\cdots,n_d}U_1^{n_1}U_2^{n_2}\cdots,
U_d^{n_d}.\end{align*} The noncommutative torus
$\mathbb{T}^d_\theta$ equipped with a complex structure $\Omega$
is called a noncommutative complex torus and is denoted by
$\mathbb{T}^d_{\theta,\Omega}$ and the algebra of smooth functions
on $\mathbb{T}^d_{\theta,\Omega}$ is denoted by
$A^d_{\theta,\Omega}$.

Let $\mathcal{E}$ be a finitely generated projective left
$A_\theta^d$-module. A connection $\nabla$ on $\mathcal{E}$ is a
linear map
\[\nabla:\mathcal{E}\lto \mathcal{E}\otimes L^*\] such that, for $x\in
L$,
\begin{align*}
\nabla_x(u\cdot e)=u\cdot\nabla_xe+\delta_x(u)\cdot e, \ \ e\in
\mathcal{E}, \ \ u\in A^d_\theta.
\end{align*}
For the integral basis $\{\lambda_i\}$ for $L$, we set
$\nabla_{\lambda_i}=\nabla_i$. The curvature of the connection
$\nabla$ is constant if
\[[\nabla_i,\nabla_j]=2\pi
iF_{ij}\cdot 1, \ \ F_{ij}\in\mathbb{R}.\] A complex (holomorphic)
structure on a right $A^d_\theta$-module $\mathcal{E}$ (compatible
with the complex structure on $A^d_{\theta,\Omega}$) is a
collection of $\mathbb{C}$-linear operators
$\overline{\nabla}_1,\cdots, \overline{\nabla}_g$ on $\mathcal{E}$
satisfying
\begin{enumerate}
\item $\overline{\nabla}_\alpha(u\cdot
e)=u\cdot\overline{\nabla}_\alpha e+\delta_{\Omega,\alpha}(u)\cdot
e$.\item $[\overline{\nabla}_\alpha,\overline{\nabla}_\beta]=0.$
\end{enumerate}
A projective left $A^d_\theta$-module $\mathcal{E}$ equipped with
holomorphic structure is called a holomorphic bundle. A vector
$\phi\in \mathcal{E}$ is called a holomorphic vector if
$\overline{\nabla}\phi=0$. We will see in the next section that
the condition 2 above assures the existence of holomorphic vectors
in $\mathcal{E}$.

Since $\mathbb{T}^d_\theta$ is a deformation quantization of the
commutative torus $\mathbb{T}^d$, a bundle over
$\mathbb{T}^d_\theta$ may be understood as a deformation of a
bundle over $\mathbb{T}^d$. In fact, the vector bundles over
$\mathbb{T}^d$ are classified by the $K$-theory of $\mathbb{T}^d$
and there is a natural isomorphism $\text{ch
}:K^*(\mathbb{T}^d)\lto H^*(\mathbb{T}^d,\mathbb{Z})\otimes
\mathbb{Q}$. Similarly, finitely generated projective
$A_\theta^d$-modules are classified by $K_0(A_\theta^d)$ and the
Chern character takes values in $H^*(\mathbb{T}^d,\mathbb{R})$.
The targets of the both Chern characters are related by the
deformation parameter $\theta\in\wedge^2L^*$. The relation is
summarized by the following diagram:
\[\begin{CD}K^0(\mathbb{T}^d)@>\text{ch}>>H^{\text{even}}(\mathbb{T}^d,\mathbb{Z})\otimes\mathbb{Q}\\
& & @V e^{i(\theta)}VV\\
K_0(A_\theta^d)@>\text{Ch}>>H^{\text{even}}(\mathbb{T}^d,\mathbb{R})
\end{CD}\]
where $i(\theta)$ denotes the contraction with 2-vector $\theta$.
Thus, for a given vector bundle $E$ over $\mathbb{T}^d$, one can
construct an $A_\theta^d$-module $\mathcal{E}$ such that
\[\text{Ch}(\mathcal{E})=e^{i(\theta)}\text{ch}(E)\] and such a
module $\mathcal{E}$ will be called a deformation of $E$ when the
zeroth component of $\text{Ch}(\mathcal{E})$ is strictly positive.
We note that such a deformation of a bundle may not be unique.
Now, in terms of cohomology, the deformation of $\mathbb{T}^d$ to
$\mathbb{T}^d_\theta$ is related with the deformation of
$H^*(\mathbb{T}^d,\mathbb{Z})$ to $H^*(\mathbb{T}^d,\mathbb{R})$
and the condition
$[\overline{\nabla}_\alpha,\overline{\nabla}_\beta]=0$ can be
understood as the existence of a cohomology class in
$H^{1,1}(\mathbb{T}^d)\cap H^2(\mathbb{T}^d,\mathbb{R})$. Finally,
for an abelian variety $(\mathbb{T}^d,\mathsf{L})$, we can
construct a pair $(\mathbb{T}^d_{\theta,\Omega},\mathcal{L})$,
where $\mathcal{L}$ is a deformation of the line bundle
$\mathsf{L}$. Such a pair
$(\mathbb{T}^d_{\theta,\Omega},\mathcal{L})$ will be referred as a
noncommutative abelian variety.

\section{Holomorphic structures on $\mathbb{T}^4_\theta$}
In this section, we apply the general discussion, given in Section
2, to the four dimensional case. We will construct a bundle over
$\mathbb{T}^4_\theta$ which is a deformation of a projectively
flat bundle over $\mathbb{T}^4$ and we will find an explicit
formula of holomorphic vectors in the deformed bundle. Finally, we
discuss how Riemann conditions correspond to the existence of
holomorphic vectors.

Let $E$ be a projectively flat $\text{U}(n)$ bundle on the complex
torus $\mathbb{T}^4=V/\Lambda$, here we follow the same notations
as in Section 2 adapted to the four dimensional case. The bundle
$E$ carries a Hermitian connection $\nabla^E$ whose curvature form
is given by
\[R_{\nabla^E}=\lambda\cdot \text{Id}_E\] where $\lambda$ is a
complex 2-form on $\mathbb{T}^4$ and $\text{Id}_E$ denotes the
identity endomorphism of $E$. The first and the second Chern
classes are given as follows:
\begin{align*}
c_1(E)&=\frac{i}{2\pi}\text{Tr
}R_{\nabla^E}=\frac{i}{2\pi}\lambda\cdot \text{rank
}E=\frac{i}{2\pi }\lambda n\\
c_2(E)&=-\frac{1}{8\pi^2}(\text{Tr }(R_{\nabla^E}\wedge
R_{\nabla^E})-\text{Tr }R_{\nabla^E}\wedge  \text{Tr
}R_{\nabla^E}=-\frac{1}{4\pi^2}\frac{n(n-1)}{2}\lambda^2\\
\frac{1}{2}c_1^2(E)&-c_2(E)=-\frac{1}{8\pi^2}n\lambda^2.
\end{align*}
Note that the Chern character of $E$ is given by
\begin{align*}
\text{ch}(E)&=\text{rank}(E)+c_1(E)+(\frac{1}{2}c_1^2(E)-c_2(E))
\in H^{\text{even}}(\mathbb{T}^4,\mathbb{Q}).
\end{align*}
The first Chern class $c_1(E)$ is an  integral, invariant 2-form
on $\mathbb{T}^4$ and we can find an integral basis
$\lambda_1,\lambda_2,\lambda_3,\lambda_4$ for $\Lambda$ such that
\[c_1(E)=\frac{1}{2}(m_1dx_{13}+m_2dx_{24}),\]
where $dx_{\mu\nu}=dx_\mu\wedge dx_\nu$. Then we have
\[\text{ch}(E)=n+\frac{1}{2}(m_1dx_{13}+m_2dx_{24})+\frac{m_1m_2}{n}
dx_{1234}.\] We take a basis for the complex vector space $V$ the
vectors
\[e_\alpha=m_\alpha^{-1}\lambda_\alpha, \ \ \alpha=1,2.\] Then the period matrix
$\mho$ for $\Lambda$ in $V$ will  be of the form
\[\mho=\begin{pmatrix}
m_1&0&\mathcal{Z}_{11}&\mathcal{Z}_{12}\\0&m_2&\mathcal{Z}_{21}&\mathcal{Z}_{22}\end{pmatrix}
:=\begin{pmatrix}\Delta_m&\mathcal{Z}\end{pmatrix}\] and by a
suitable basis change, we can take the dual period matrix $\Omega$
as in the following form
\[\Omega=\begin{pmatrix}\frac{1}{m_1}&0&Z_{11}&Z_{12}\\0&\frac{1}{m_2}&Z_{21}&Z_{22}\end{pmatrix}:=
\begin{pmatrix}\Delta_m^{-1}&Z\end{pmatrix},\] where $Z=-\overline{\mathcal{Z}}$ ({\it cf.} \cite{GH}).
Let us denote the corresponding integral, skew-symmetric $4\times
4$ matrix $q$ by
\[q=\begin{pmatrix}0&\Delta_m\\-\Delta_m&0\end{pmatrix}.\]

On the other hand, the bundle $E$ can be deformed to a projective
module $\mathcal{E}$ over the noncommutative torus
$\mathbb{T}^4_\theta$ whose Chern character is given in the
following form
\[\text{Ch}(\mathcal{E})=e^{i(\theta)}\text{ch}(E)\in
H^{\text{even}}(\mathbb{T}^4,\mathbb{R}),\] where $i(\theta)$
denotes the contraction with $\theta\in\wedge^2L^*$. More
explicitly, we have
\begin{align*}
\text{Ch}(\mathcal{E})=&(n+\frac{1}{2}\text{Tr}(q\theta)+
\frac{m_1m_2}{n}\text{Pf}(\theta))+\frac{1}{2}(q+*\theta
\frac{m_1m_2}{n})_{ij}dx_{ij}+\frac{m_1m_2}{n}dx_{1234},
\end{align*}
where
$\text{Pf}(\theta))=\theta_{13}\theta_{24}-\theta_{12}\theta_{34}-\theta_{14}\theta_{23}$
is the Paffian of  $\theta$. If there is a constant curvature
connection in $\mathcal{E}$, then the corresponding curvature up
to a factor $2\pi i$ can be identified with matrix
\[F=2\pi i\frac{q+*\theta
\frac{m_1m_2}{n}}{n+\frac{1}{2}\text{Tr}(q\theta)+
\frac{m_1m_2}{n}\text{Pf}(\theta)}=2\pi i\frac{q+*\theta
\frac{m_1m_2}{n}}{\text{dim }\mathcal{E}}.\]

Let $\gamma=(\gamma_{ij})=\psi-\theta$, where
\begin{align}\label{rational}\psi=(\psi_{ij})=\begin{pmatrix}0&0&\frac{n}{m_1}&0\\
0&0&0&\frac{n}{m_2}\\
-\frac{n}{m_1}&0&0&0\\
0&-\frac{n}{m_2}&0&0
\end{pmatrix}=-nq^{-1}.\end{align}
Then
\begin{align*}
\text{Pf}(\gamma)&=\text{Pf}(\psi-\theta)\\
&=(\frac{n}{m_1}-\theta_{13})(\frac{n}{m_2}-\theta_{24})-\theta_{12}\theta_{34}-\theta_{14}\theta_{23}\\
&=\frac{n}{m_1m_2}\left(n+\frac{1}{2}\text{Tr}(q\theta)+
\frac{m_1m_2}{n}\text{Pf}(\theta)\right)
\end{align*}
and since $\gamma_{ij}=\psi_{ij}-\theta_{ij}$, we have
\[\gamma^{-1}=\frac{1}{\text{Pf}(\gamma)}\begin{pmatrix}0&\gamma_{34}&-\gamma_{24}&\gamma_{23}\\
-\gamma_{34}&0&\gamma_{14}&-\gamma_{13}\\
\gamma_{24}&-\gamma_{14}&0&\gamma_{12}\\
-\gamma_{23}&\gamma_{13}&-\gamma_{12}&0\end{pmatrix}=-\frac{q+*\theta
\frac{m_1m_2}{n}}{n+\frac{1}{2}\text{Tr}(q\theta)+
\frac{m_1m_2}{n}\text{Pf}(\theta)}.\] Thus
\[F=\frac{2\pi}{i}\gamma^{-1}.\]

Note that if, for $\theta$ not rational, the zeroth component of
the Chern character is strictly positive, then the gauge bundle
$\mathcal{E}$ belongs to the positive cone of $K_0(A_\theta^4)$
and it can be written as a direct sum of the form
$\mathcal{S}(\mathbb{R}^2\times F)$ (\cite{Ri1}), where $F$ is a
finite abelian group. Thus by assuming $\text{dim }\mathcal{E}>0$,
let $\mathcal{E}=\mathcal{S}(\mathbb{R}^2\times F)$. The
projective flat bundle $E$ over $\mathbb{T}^4$ can be specified by
operators $W_1,W_2,W_3,W_4$ acting on $C(\mathbb{Z}_{m_1}\times
\mathbb{Z}_{m_2})=\mathbb{C}^{m_1}\otimes \mathbb{C}^{m_2}$:
\begin{align*}
W_1f(k_1,k_2)&=f(k_1-n,k_2) \ \ \ \ \ \ \ \ \ \ \ \ \ \ \ \ \ \ \ \  W_2f(k_1,k_2)=f(k_1,k_2-n)\\
W_3f(k_1,k_2)&=\exp(-2\pi i\frac{k_1}{m_1})f(k_1,k_2) \ \ \ \ \ \
\  W_4f(k_1,k_2)=\exp(-2\pi i\frac{k_2}{m_2})f(k_1,k_2).
\end{align*}
The operators obey the commutation relation
\[W_iW_j=\exp(2\pi i\psi_{ij})W_jW_i.\]
Let us consider an embedding of $\Lambda$ into
$\mathbb{R}^2\times(\mathbb{R}^2)^*$ in the sense of \cite{Ri1}.
Such an embedding map can be represented by a non-singular matrix
$T=(T_{ij})$ such that
\[(\wedge^2 T^*)(e_3\wedge e_2+e_4\wedge e_1)=-\gamma\in \wedge^2
\mathbb{R}^2.\] Associated to the embedding $T$, the left module
action on $\mathcal{S}(\mathbb{R}^2)$ is defined by the Heisenberg
representation of $\Lambda$ in $\mathcal{S}(\mathbb{R}^2)$:
\[(V_if)(r,s):=(V_{e_i}f)(r,s)=\exp[2\pi
i(rT_{3i}+sT_{4i})]f(r+T_{1i},s+T_{2i}),\] which obeys the
commutation relations:
\[V_iV_j=\exp(-2\pi i\gamma_{ij})V_jV_i,\] and
\[\gamma_{ij}=\left|\begin{matrix}T_{1i}&T_{1j}\\T_{3i}&T_{3j}\end{matrix}\right|+
\left|\begin{matrix}T_{2i}&T_{2j}\\T_{4i}&T_{4j}\end{matrix}\right|.\]
Let us consider the operator $U_i=V_i\otimes W_i$ acting on
$\mathcal{E}=\mathcal{S}(\mathbb{R}^2\times
\mathbb{Z}_{m_1}\times\mathbb{Z}_{m_2})$. The operators $U_i$'s
hold the commutation relation
\[U_iU_j=\exp(2\pi i\theta_{ij})U_jU_i\]
and hence it defines a left $A_\theta^4$-module action on
$\mathcal{E}$. Analogously, we may define a constant curvature
connection on $\mathcal{E}$ as follows: for
$(r,s)\in\mathbb{R}^2$,
\begin{equation}\label{connection}
(\nabla_if)(r,s)=2\pi irA_{i1}f(r,s)+2\pi
isA_{i2}f(r,s)-A_{i3}\frac{\partial f}{\partial
r}(r,s)-A_{i4}\frac{\partial f}{\partial s}(r,s).
\end{equation}
It is easy to compute the curvature of $\nabla$ and it takes of
the form
\[F_{ij}=[\nabla_i,\nabla_j]=2\pi
i\left(\left|\begin{matrix}A_{i1}&A_{i3}\\A_{j1}&A_{j3}\end{matrix}\right|
+\left|\begin{matrix}A_{i2}&A_{i4}\\A_{j2}&A_{j4}\end{matrix}\right|\right).\]
Thus if $A=(A_{ij})=(T^{-1})^*$, then we get
\[F_{ij}=-\frac{2\pi i}{\text{Pf}(\gamma)}*\gamma_{ij}\] and hence
the curvature matrix of $\nabla$ is identified with $-2\pi
i\gamma^{-1}$.

Let us consider the noncommutative complex torus
$\mathbb{T}^4_{\theta,\Omega}$, where  the complex structure on
$\mathbb{T}^4_\theta$ is specified by the dual period matrix
$\Omega=\begin{pmatrix}\Delta_m^{-1}&Z\end{pmatrix}$ of $\Lambda$
in $V$. The compatible holomorphic structure on the deformed
bundle or the finitely generated projective left
$A_\theta^4$-module $\mathcal{E}$ is determined by the following
defining equation
\begin{equation}\label{defining}
[\overline{\nabla}_\alpha,\overline{\nabla}_\beta]=0, \ \ \
\alpha, \beta=1,2
\end{equation} where $\overline{\nabla}=\Omega \nabla$, and hence
\[\begin{pmatrix}\overline{\nabla}_1\\
\overline{\nabla}_2\end{pmatrix}=\begin{pmatrix}
\frac{1}{m_1}&0&Z_{11}&Z_{12}\\0&\frac{1}{m_2}&Z_{21}&Z_{22}\end{pmatrix}
\begin{pmatrix}
\nabla_1\\\nabla_2\\\nabla_3\\\nabla_4\end{pmatrix}=
\begin{pmatrix}
\frac{1}{m_1}\nabla_1+Z_{11}\nabla_3+Z_{12}\nabla_4\\\frac{1}{m_2}\nabla_2+Z_{21}\nabla_3+Z_{22}\nabla_4\end{pmatrix}.\]
Since the curvature matrix $F$ of $\nabla$ is $-2\pi
i\gamma^{-1}$, the defining equation (\ref{defining}) is
equivalent to the equation
\[\Omega\gamma^{-1}\Omega^t=0.\]
Also we have a useful defining equation
\begin{align}
[\overline{\nabla}_\alpha,\overline{\nabla}_\beta]&=
[\frac{1}{m_1}\nabla_1+Z_{11}\nabla_3+Z_{12}\nabla_4,\frac{1}{m_2}\nabla_2
+Z_{21}\nabla_3+Z_{22}\nabla_4]\nonumber\\
&=\frac{1}{m_1m_2}\gamma_{34}+\det
Z\gamma_{12}-\frac{1}{m_2}Z_{11}\gamma_{14}+\frac{1}{m_2}Z_{12}\gamma_{13}
-\frac{1}{m_1}Z_{21}\gamma_{24}-\frac{1}{m_1}Z_{22}\gamma_{23}\label{def2}\\
&=0\nonumber
\end{align}

We now find a holomorphic vector in $\mathcal{E}$, which is a
solution for the equation
\begin{equation}\label{eq1}
\overline{\nabla}\phi(r,s)=0,
\end{equation}
By the definition of $\overline{\nabla}$ and (\ref{connection}),
we have
\begin{equation}\label{holo}
\overline{\nabla}_\alpha f(r,s)=2\pi ir\Omega^\alpha\cdot A_1
f(r,s)+2\pi is\Omega^\alpha\cdot A_2 f(r,s)-\Omega^\alpha\cdot
A_3\frac{\partial f}{\partial r}(r,s) -\Omega^\alpha\cdot
A_4\frac{\partial f}{\partial s}(r,s)
\end{equation}
where $\Omega^\alpha$ denotes the $\alpha$-th row of $\Omega$ and
$A_i$ is the $i$-th column of the matrix $A$.

The solution of the equation (\ref{eq1}) should be of the form
\[\phi(r,s)=\exp[\pi
i\begin{pmatrix}r&s\end{pmatrix}H\begin{pmatrix}r\\s\end{pmatrix}],\]
where $H=\begin{pmatrix}H_{11}&H_{12}\\H_{21}&H_{22}\end{pmatrix}$
is a complex $2\times 2$ matrix. If $H$ is not symmetric, then the
equation (\ref{eq1}) does not have a solution, and
$\phi\in\mathcal{S}(\mathbb{R}^2)$ only when $\text{Im }H>0$. Thus
we assume that $H$ is symmetric and $\text{Im }H>0$. Then
\[\phi(r,s)=\exp[\pi i(r^2H_{11}+s^2H_{22}+2rsH_{12})]\] and hence
\begin{align*}
\frac{\partial\phi}{\partial r}(r,s)&=2\pi
i(rH_{11}+sH_{12})\phi(r,s)\\
\frac{\partial\phi}{\partial s}(r,s)&=2\pi
i(rH_{12}+sH_{22})\phi(r,s)
\end{align*}
Applying (\ref{holo}) to the equation (\ref{eq1}), we have four
equations with three unknowns $H_{11}$, $H_{12}$, and $H_{22}$:
\begin{align*}
(\Omega^1\cdot A_3)H_{11}+(\Omega^1\cdot A_4)
H_{12}&=\Omega^1\cdot
A_1\\
(\Omega^1\cdot A_3)H_{12}+(\Omega^1\cdot A_4)
H_{22}&=\Omega^1\cdot
A_2\\
(\Omega^2\cdot A_3)H_{11}+(\Omega^2\cdot A_4)
H_{12}&=\Omega^2\cdot
A_1\\
(\Omega^2\cdot A_3)H_{12}+(\Omega^2\cdot A_4)
H_{22}&=\Omega^2\cdot A_2
\end{align*}
Such a linear system of equations will have a solution if
\[\left|\begin{matrix}
\Omega^1\cdot A_3&0&\Omega^1\cdot A_4&\Omega^1\cdot A_1\\
0&\Omega^1\cdot A_4&\Omega^1\cdot A_3&\Omega^1\cdot A_2\\
\Omega^2\cdot A_3&0&\Omega^2\cdot A_4&\Omega^2\cdot A_1\\
0&\Omega^2\cdot A_4&\Omega^2\cdot A_3&\Omega^2\cdot A_2
\end{matrix}\right|=0.\]
In the above, the determinant is computed to
\begin{align*}
& \left\{ (\Omega^1\cdot A_1)(\Omega^2\cdot A_3)-(\Omega^2\cdot
A_1)(\Omega^1\cdot A_3)+(\Omega^1\cdot A_2)(\Omega^2\cdot
A_4)-(\Omega^2\cdot A_2)(\Omega^1\cdot A_4) \right\}\\&\times
\left\{ (\Omega^1\cdot A_3)(\Omega^2\cdot A_4)-(\Omega^2\cdot
A_3)(\Omega^1\cdot A_4) \right\}.
\end{align*}
We first compute the following:
\begin{align}
(\Omega^1&\cdot A_i)(\Omega^2\cdot A_j)-(\Omega^2\cdot
A_i)(\Omega^1\cdot
A_j)\label{ceq}\\
&=\frac{1}{m_1m_2}\left|\begin{matrix}A_{1i}&A_{1j}\\A_{2i}&A_{2j}\end{matrix}\right|
-\frac{1}{m_2}Z_{11}\left|\begin{matrix}A_{2i}&A_{2j}\\A_{3i}&A_{3j}\end{matrix}\right|
-\frac{1}{m_2}Z_{12}\left|\begin{matrix}A_{2i}&A_{2j}\\A_{4i}&A_{4j}\end{matrix}\right|\nonumber\\
& \ \
+\frac{1}{m_1}Z_{21}\left|\begin{matrix}A_{1i}&A_{1j}\\A_{3i}&A_{3j}\end{matrix}\right|
+\frac{1}{m_1}Z_{22}\left|\begin{matrix}A_{1i}&A_{1j}\\A_{4i}&A_{4j}\end{matrix}\right|
+\det Z
\left|\begin{matrix}A_{3i}&A_{3j}\\A_{4i}&A_{4j}\end{matrix}\right|.\nonumber
\end{align}
In particular, if either $m_1=m_2$ or $Z$ is symmetric, we can
write (\ref{ceq}) in the following form:
\[(\Omega^1\cdot A_i)(\Omega^2\cdot A_j)-(\Omega^2\cdot
A_i)(\Omega^1\cdot A_j)=\left|\begin{matrix}
-Z_{11}&-Z_{12}&A_{1i}&A_{1j}\\
-Z_{21}&-Z_{22}&A_{2i}&A_{2j}\\
\frac{1}{m_1}&0&A_{3i}&A_{3j}\\
0&\frac{1}{m_2}&A_{4i}&A_{4j}
\end{matrix}\right|.\]
Using the identity (\ref{ceq}), we now have
\begin{align*}
&\left\{ (\Omega^1\cdot A_1)(\Omega^2\cdot A_3)-(\Omega^2\cdot
A_1)(\Omega^1\cdot A_3)+(\Omega^1\cdot A_2)(\Omega^2\cdot
A_4)-(\Omega^2\cdot A_2)(\Omega^1\cdot A_4) \right\}\\
&=\frac{1}{m_1m_2}\gamma_{34}+\det
Z\gamma_{12}-\frac{1}{m_2}Z_{11}\gamma_{14}+\frac{1}{m_2}Z_{12}\gamma_{13}
-\frac{1}{m_1}Z_{21}\gamma_{24}-\frac{1}{m_1}Z_{22}\gamma_{23}\\
&=0 \ \ \ \text{ by (\ref{def2})}.
\end{align*}
Thus the equation (\ref{eq1}) has a solution. As a conclusion, the
condition $[\overline{\nabla}_\alpha,\overline{\nabla}_\beta]=0$
implies that holomorphic vectors in $\mathcal{E}$ do exist. By
assuming
\[(\Omega^1\cdot A_3)(\Omega^2\cdot A_4)-(\Omega^2\cdot
A_3)(\Omega^1\cdot A_4) \ne 0\] we have the solution for the
equation (\ref{eq1}) as follows:

\begin{align*}
H_{11}&=\frac{(\Omega^1\cdot A_1)(\Omega^2\cdot
A_4)-(\Omega^2\cdot A_1)(\Omega^1\cdot A_4)}{(\Omega^1\cdot
A_3)(\Omega^2\cdot A_4)-(\Omega^2\cdot A_3)(\Omega^1\cdot A_4)}\\
\\ H_{12}=H_{21}&=\frac{(\Omega^1\cdot A_3)(\Omega^2\cdot
A_1)-(\Omega^2\cdot A_3)(\Omega^1\cdot A_1)}{(\Omega^1\cdot
A_3)(\Omega^2\cdot A_4)-(\Omega^2\cdot A_3)(\Omega^1\cdot A_4)}\\
\\ H_{22}&=\frac{(\Omega^1\cdot A_2)(\Omega^2\cdot
A_3)-(\Omega^2\cdot A_2)(\Omega^1\cdot A_3)}{(\Omega^1\cdot
A_3)(\Omega^2\cdot A_4)-(\Omega^2\cdot A_3)(\Omega^1\cdot A_4)}.
\end{align*}

We have shown that the defining equation (\ref{def2}) assures the
existence of holomorphic vectors in a holomorphic bundle over
$\mathbb{T}^4_{\theta,\Omega}$. In what follows, we discuss more
consequences of the equation (\ref{def2}).

First note that in terms of the rational matrix $\psi=-nq^{-1}$
defined in (\ref{rational}) we can state the the ordinary Riemann
condition I as follows: The commutative complex torus
$\mathbb{T}^4=V/\Lambda$ is an abelian variety if and only if
\begin{equation}\label{comrie}
\Omega\psi^{-1}\Omega^t=0, \ \ \ \ \
i\Omega\psi^{-1}\overline{\Omega}^t>0
\end{equation}
On the other hand, the equation (\ref{def2}) is equivalent to
\begin{equation}\label{ncriemann}
\Omega\gamma^{-1}\Omega^t=\Omega(\psi-\theta)^{-1}\Omega^t=0
\end{equation}
and the equation determines a holomorphic structure on a bundle
over $\mathbb{T}^4_{\theta,\Omega}$. As discussed in Section 2,
the deformation of commutative torus $\mathbb{T}^4$ to  a
noncommutative torus $\mathbb{T}^4_\theta$ corresponds to the
cohomological deformation from $H^*(\mathbb{T}^4,\mathbb{Z})$ to
$H^*(\mathbb{T}^4,\mathbb{R})=\wedge^\bullet L^*$. Since the
construction of the equation (\ref{ncriemann}) reflects the
cohomological deformation, we might call the equation
(\ref{ncriemann}) as {\it noncommutative Riemann condition} under
the condition that the corresponding solution to the equation
$\overline{\nabla}\phi=0$ is in $\mathcal{S}(\mathbb{R}^2)$.

Topologically, any commutative torus $\mathbb{T}^4$ can be
deformed to a noncommutative torus $\mathbb{T}^4_\theta$ with the
deformation parameter $\theta\in \wedge^2L^*$. Along with this,
one can ask a following question: For any deformation parameter
$\theta$, can every abelian variety be deformed to a
noncommutative abelian variety? More precisely, for an abelian
variety $\mathbb{T}^4=V/\Lambda$ equipped with a line bundle
$\mathsf{L}$ such that $c_1(\mathsf{L})=m_1dx_{13}+m_2dx_{24}$,
the question is to construct a holomorphic bundle $\mathcal{L}$
over $\mathbb{T}^4_\theta$ which is a deformation of $\mathsf{L}$
in the sense of the definition given in Section 2. In what
follows, we discuss that the answer is negative.  Let us consider
a simple example where the deformation parameter is given in the
following form:
\[\theta=\begin{pmatrix}0&0&\theta_{13}&0\\0&0&0&\theta_{24}\\
-\theta_{13}&0&0&0\\0&-\theta_{24}&0&0\end{pmatrix}\] and thus
\[\gamma=\begin{pmatrix}0&0&\gamma_{13}&0\\0&0&0&\gamma_{24}\\
-\gamma_{13}&0&0&0\\0&-\gamma_{24}&0&0\end{pmatrix}=
\begin{pmatrix}0&0&\frac{1}{m_1}-\theta_{13}&0\\0&0&0&\frac{1}{m_2}-\theta_{24}\\
-\frac{1}{m_1}+\theta_{13}&0&0&0\\0&-\frac{1}{m_2}+\theta_{24}&0&0\end{pmatrix}.\]
In this case, the defining equation or the noncommutative Riemann
condition is
\begin{equation}\label{simple}
[\overline{\nabla}_1,\overline{\nabla}_2]=\frac{1}{m_2}Z_{12}\gamma_{13}-\frac{1}{m_1}Z_{21}\gamma_{24}=0.
\end{equation} If the complex torus $\mathbb{T}^4=V/\Lambda$ is an abelian
variety with the integral skew-symmetric matrix $q$, $n=1$, then
the complex matrix
$Z=\begin{pmatrix}Z_{11}&Z_{12}\\Z_{21}&Z_{22}\end{pmatrix}$ is
symmetric and the equation (\ref{simple}) becomes
\[Z_{12}(\frac{\gamma_{13}}{m_2}-\frac{\gamma_{24}}{m_1})=0.\] Since
$Z_{12}=Z_{21}$ is arbitrary, we have
\begin{equation}\label{nega}
\frac{\gamma_{13}}{m_2}=\frac{\gamma_{24}}{m_1}\text{ \ \ \ or \ \
\ }\frac{\theta_{13}}{m_2}=\frac{\theta_{24}}{m_1}.\end{equation}
Thus we see that the deformation parameter $\theta$ should be
restricted. In other words, for an arbitrary $\theta$, one cannot
have a noncommutative abelian variety
$(\mathbb{T}^4_{\theta,\Omega},\mathcal{L})$ which is a
deformation of $(\mathbb{T}^4,\mathsf{L})$ and
$(\mathbb{T}^4_{\theta,\Omega},\mathcal{L})$ is defined only when
$\theta$ holds the equation (\ref{nega}).

As discussed in Section 2, the deformation of $\mathbb{T}^4$ to
$\mathbb{T}^4_\theta$ corresponds to the cohomological deformation
$H^*(\mathbb{T}^4,\mathbb{Z})$ to  $H^*(\mathbb{T}^4,\mathbb{R})$.
On a commutative complex torus $\mathbb{T}^4$, the Riemann
conditions are obtained by comparing the rational structure and
the complex structure on $H^*(\mathbb{T}^4)$. Thus, associated to
the noncommutative Riemann conditions, it is natural to consider
the relation between the real structure and the complex structure
on $H^*(\mathbb{T}^4)$. Note that the cohomological deformation
can be seen in the curvature matrices. Let us consider the
constant curvature connection $\nabla$ given in (\ref{connection})
whose curvature matrix is $\gamma^{-1}$. If we assume the
Euclidean metric for $L=\mathbb{R}^4$, we can use the orthogonal
transformation to change  matrix $\gamma$ into the standard form:
\[\tilde\gamma=\begin{pmatrix}
0&0&\gamma_1&0\\0&0&0&\gamma_2\\-\gamma_1&0&0&0\\0&-\gamma_2&0&0\end{pmatrix}
=P\gamma^{-1}P^t, \ \ \ P\in \text{O}(4).\] Note that the
connection $\nabla$ is unitary equivalent to the connection
$\widetilde{\nabla}$ whose curvature is $\tilde\gamma^{-1}$. In
terms of cohomologies, for any invariant, real 2-form
\[\sum_{i=1}^4\gamma_{ij}dx_i\wedge dx_j,\] one can
find an orthonormal basis $\xi_1,\cdots \xi_4$ for $L$ such that
\[\omega=\frac{1}{2}(\gamma_1^{-1}dx_{13}+\gamma_2^{-1}dx_{24}).\]
We take a basis for $V$ the vectors
\[\zeta_\alpha=\gamma_\alpha^{-1}\xi_\alpha, \ \ \alpha=1,2.\]
Then the corresponding complex structure will be of the form
\[\widetilde{\Omega}=\begin{pmatrix}\gamma_1&0&\widetilde{Z}_{11}&\widetilde{Z}_{12}\\
0&\gamma_2&\widetilde{Z}_{21}&\widetilde{Z}_{22}\end{pmatrix}.\]
We use the complex structure to define noncommutative complex
torus $\mathbb{T}^4_{\theta,\widetilde{\Omega}}$ and a holomorphic
structure on a bundle over $T^4_\theta$ equipped with the constant
curvature connection $\widetilde\nabla$. By the same argument as
given above, we have the defining equation (\ref{defining}) in the
following form
\[\widetilde{\Omega}\tilde\gamma^{-1}\widetilde{\Omega}^t=0.\]
Since
\begin{align*}
&\begin{pmatrix}\gamma_1&0&\widetilde{Z}_{11}&\widetilde{Z}_{12}\\
0&\gamma_2&\widetilde{Z}_{21}&\widetilde{Z}_{22}\end{pmatrix}
\begin{pmatrix}
0&0&-\gamma_1^{-1}&0\\0&0&0&-\gamma_2^{-1}\\\gamma_1^{-1}&0&0&0\\0&\gamma_2^{-1}&0&0\end{pmatrix}
\begin{pmatrix}\gamma_1&0\\0&\gamma_2\\\widetilde{Z}_{11}&\widetilde{Z}_{21}
\\\widetilde{Z}_{12}&\widetilde{Z}_{22}\end{pmatrix}\\
&=\begin{pmatrix}0&\widetilde{Z}_{12}-\widetilde{Z}_{21}\\\widetilde{Z}_{21}-\widetilde{Z}_{12}&0\end{pmatrix},
\end{align*}
we see that the matrix
$\widetilde{Z}=\begin{pmatrix}\widetilde{Z}_{11}&\widetilde{Z}_{12}\\
\widetilde{Z}_{21}&\widetilde{Z}_{22}\end{pmatrix}$
is symmetric. In this case. we can find a holomorphic vector in a
simple form:
\[\phi(r,s)=\exp[\pi
i\begin{pmatrix}\gamma_1r&\gamma_2s\end{pmatrix}\widetilde{Z}
\begin{pmatrix}\gamma_1r\\\gamma_2s\end{pmatrix}].\]
This can be obtained from the form of solution obtained above by
taking suitable entries of the matrix $A$ as given in \cite{KL}.
Also, we see that the solution is in $\mathcal{S}(\mathbb{R}^2)$
only when the matrix $\text{Im }\widetilde{Z}$ is negative
definite. In \cite{Sch01}, this kind of solution has been obtained
in case of $\gamma_1=\gamma_2=1$ and hence it is easily seen from
our solution how deformation parameter is involved in holomorphic
vectors.

\section{Conclusion}
In this paper, we found an explicit solution for holomorphic
connections on a four dimensional noncommutative complex torus. We
also related the definition of a holomorphic structure for a
bundle over a noncommutative complex torus to the ordinary Riemann
conditions. In doing so, we defined a deformation of abelian
variety along the Chern character deformation and we found that
deformation parameter should be restricted to define a
noncommutative abelian variety. Finally, we studied a
noncommutative variation of Kodaira's embedding theorem. By using
the real cohomology class instead integral ones, we obtain a
noncommutative version of Riemann conditions. We expect all these
results should be a first step to the study of the mirror symmetry
for noncommutative complex torus or noncommutative abelian
varieties. We will come to this in forthcoming paper.

\begin{center}
{\large \bf Acknowledgments} \\
\end{center}
The second author is supported by KRF-2003.

\end{document}